\documentclass[article,aps,twocolumn]{revtex4-1}  

\usepackage{amsmath}    
\usepackage{graphicx}   

\DeclareMathOperator{\sinc}{sinc}

\newcommand\Expect[1]{\mathrm{E}\left[#1\right]}

\begin{document}

\title{Quantum modulation against electromagnetic interference}
\author{Juan Carlos Garcia-Escartin}
 \email{juagar@tel.uva.es}  
\affiliation{Universidad de Valladolid, Dpto. Teor\'ia de la Se\~{n}al e Ing. Telem\'atica, Paseo Bel\'en n$^o$ 15, 47011 Valladolid, Spain}
\date{\today}

\begin{abstract}
Periodic signals in electrical and electronic equipment can cause interference in nearby devices. Randomized modulation of those signals spreads their energy through the frequency spectrum and can help to mitigate electromagnetic interference problems. The inherently random nature of quantum phenomena makes them a good control signal. I present a quantum modulation method based on the random statistics of quantum light. The paper describes pulse width modulation schemes where a Poissonian light source acts as a random control that spreads the energy of the potential interfering signals. I give an example application for switching-mode power supplies and comment the further possibilities of the method.
\end{abstract}
\maketitle

\section{Electromagnetic compatibility and random modulation}
The widespread adoption of electrical and electronic devices makes electronic interference a growing concern. Electromagnetic compatibility studies how the signals inside a particular piece of equipment can create interference in nearby devices and how to avoid it. The signals that perform a useful function can also inadvertently feed unintentional antennas (and produce radiated emissions) or send residual noise into the common electrical line (conducted emissions) \cite{Pau06}. There are different ways to control interference. They include electromagnetic shielding, introducing filters, careful grounding or distributing the wires in configurations that reduce the overall emissions. There is no one-size-fits-all solution and most devices use a combination of methods. 

One particularly vexing source of electromagnetic interference (EMI) are periodic signals. Signals with a period $T$ concentrate their energy in the harmonics $kf_0$ of the fundamental frequency $f_0=\frac{1}{T}$ with integer $k$ and can reach high powers in a narrow frequency band. Spread spectrum solutions offer a possible mitigation method. In spread spectrum methods the periodic signal is modulated so that its energy is stretched through the frequency spectrum. The total energy of the potentially interfering signal is the same, but at any single frequency or any small frequency band, the energy is lower. There is some debate on whether spread spectrum modulation helps when the receiving end acts as a wideband receiver. Nevertheless, coupling mechanisms that take the energy from one device to another are usually efficient only in a limited frequency band. Depending on the frequency, wires acting as antennas are more or less efficient and conducted emissions see different ground paths. In certain situations, modulation has proven to be effective even to reduce wideband interference \cite{SS01,HOF03}.

The value of spread spectrum methods is implicitly recognized in EMI industry standards. These standards limit the amount of electromagnetic interference an electronic device can emit before it can be commercially sold. The most important standards \cite{CFR15,EU04,EN301} impose limits on the maximum power at any frequency from a certain range (or in a limited frequency band, due to the measurement method). For that reason, the industry has widely adopted spread spectrum methods, if only to help them pass the EMI tests.

Spread spectrum methods are particularly effective on periodic signals, as they already have a concentrated energy content. Randomized modulation schemes are a good way to spread the energy of the signal through a wider frequency band \cite{SVG95}. In them, a random-looking control signal modulates the periodic input so that the functional requirements of the input are preserved while the spectrum is wider but lower. The modulating signal can be just a periodic function of a different, lower period \cite{BSO05} or a signal coming from pseudo-random number generators \cite{TWP07}, chaotic oscillators \cite{TNS03} or Markov-chain processes \cite{SVP97}. 

In this paper, I propose a modulation method based on a quantum random signal. The inherently random nature of quantum phenomena makes them a natural choice as sources of randomness. Many true random number generators are based on quantum measurements. From the early systems based on the detection of radioactive decay \cite{Vin70}, there have been many proposed quantum random number generators, most of them based on quantum optics \cite{JAW00,SGG00,SR07}.

This paper studies the application of quantum randomness to spread spectrum modulation of periodic signals. Unlike in quantum random number generators, we do not need a uniform distribution and we can use just a detector with minimal post-processing.

\section{Switching-mode power supplies}
As an application example, we can take switching-mode power supplies, SMPSs. A switching-mode power supply acts as a power converter by periodically switching between on and off states. An important parameter is its duty cycle, $D$, which gives the fraction of the period $T$ in which the signal state is on. The duty cycle can be adjusted to convert between different power ratings. SMPSs have high electromagnetic emissions due to both the periodicity of the control signal, which concentrates the energy of potentially interfering signals in a few harmonics, and due to the usually high power levels of the signal, which means there is a larger interfering power. 

\begin{figure}[ht!]
\includegraphics{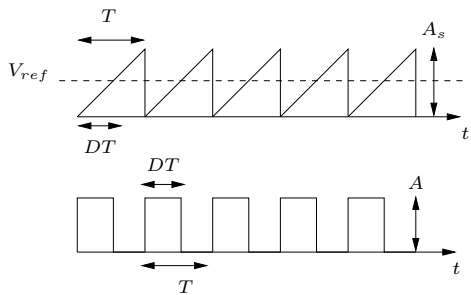}
\caption{\label{PWM} Pulse width modulation from the comparison of a sawtooth function and a reference voltage $V_{ref}$.}
\end{figure}

The control signal of an SMPS is a train of pulses of period $T$ and width $DT$. Figure \ref{PWM} shows a common way to generate such a signal using a pulse width modulation, PWM, system with a sawtooth signal $f_{saw}(t)$ and a reference voltage $V_{re\!f}$. Both signals are fed into a comparator that outputs a signal of amplitude $A$ whenever $f_{saw}(t)>V_{re\!f}$. A change in $V_{re\!f}$ changes $D$. If the sawtooth signal has a maximum amplitude $A_{s}$, a reference voltage $V_{re\!f}=DA_{s}$ selects the desired duty cycle.

The spectrum of the output PWM signal can be described by its Fourier coefficients. We have a signal that comes from the periodic repetition with a period $T$ of a basic pulse in the $[0,T]$ interval defined as
\begin{equation}
u_T(t) = \left\{
  \begin{array}{lr}
    A, & \text{if } 0<t<DT\\
    0, & \text{otherwise}
  \end{array}
\right.
\end{equation}
which has Fourier coefficients
\begin{equation}
c_k=e^{-j k \pi D}AD\sinc(kD),
\end{equation}
for any integer $k$ and with $\sinc(x)=\frac{\sin(\pi x)}{\pi x}$. The average power in one period of the signal is 
\begin{equation}
P_{av}=\frac{1}{T}\int_{t'}^{t'+T}u_T^2(t)dt=|c_0|^2+\sum_{k=1}^{\infty}2|c_k|^2.
\end{equation}
As expected from Parserval's theorem, the Fourier coefficients give us the total power and we can separate the DC component given by $c_0$ and the the power around the $k$th harmonic $2|c_k|^2$, if we consider only positive $k$s (unilateral spectrum). The zero-frequency component given by $c_0$ cannot be radiated and will hardly affect the total interference. In the rest of the paper, we will concern ourselves mostly with the rest of the harmonics. 

In spread spectrum EMI reduction, we add a random control that preserves the average duty cycle but, at the same time, introduces enough unpredictability to turn the periodic signal with concentrated power into a wide spectrum signal. A quantum modulation system will use a quantum signal as the origin of this randomness. In principle there are many quantum phenomena that can play the role of the modulating signal: radioactive decay, measurement in quantum dots or ion trap system, etc. However, quantum light is a natural first option. It can be produced and detected with efficiency and there are many technologies that can be employed.

\section{Quantum light as a source of randomness}

We can take a source of coherent light such as a laser or a LED diode in time intervals much smaller than its coherence time. If they are fed with a constant current, the resulting photons follow a Poissonian distribution \cite{Lou00}. For a current $I$ that guarantees there is an average of $\lambda$ photons per second, in an observation time $T_{ob}$, the probability of finding $n$ photons is
\begin{equation}
P(n,T_{ob})=\frac{(\lambda T_{ob})^n}{n!}e^{-\lambda T_{ob}}.
\end{equation}
The power fluctuations that come from the discrete nature of light bring an unavoidable quantum noise that can be turned to our advantage as a source of randomness. For simplicity, in the following, I will assume a Poissonian source of light. However, we can tailor the photon distribution if we modify the current \cite{KS68,WK10}. 

\section{Quantum modulation}
The photons are converted into voltage using a photodetector. We can use a PIN photodiode or a CCD camera to collect the photons that arrive in a time of detection $T_{det}$ and produce a voltage proportional to the number of photons found, $n$. To make things simpler, we can consider a system that creates a new random voltage value in each period. We consider $T_{det}=T$ and an average photon number $\lambda_T=\lambda T$ in the measured interval. We can imagine the detector is refreshed every $T$ seconds and that we use a sample-and-hold circuit that keeps for a time $T$ the voltage produced from the photons detected in a shorter time of detection or in the previous period. This covers the scenarios where the detector cannot work continuously. 

\begin{figure}[ht!]
\includegraphics{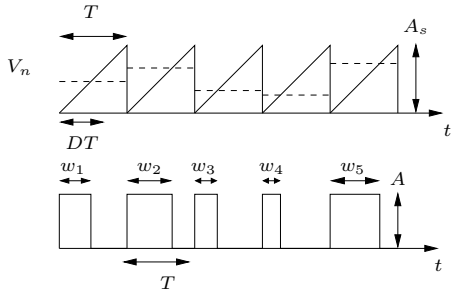}
\caption{\label{QPWM}Quantum pulse width modulation. The pulses are generated from the comparison of a sawtooth function and a random voltage $V_{n}$ proportional to the number of photons detected.}
\end{figure}

The circuit produces random voltages $V_n=g n$ proportional to the detected number of photons $n$ with a gain $g$ that depends on the detector. For the time being we suppose a noiseless detector. A quantum source of coherent light fed with a constant current can thus create a random Poissonian voltage that will serve as the modulating signal in a spread spectrum EMI reduction system. Figure \ref{QPWM} shows an example of quantum modulation in an SMPS. With quantum modulation, the output signal is a sequence of pulses of random widths, with one pulse every $T$ seconds. The power distribution in frequency of a random signal can be described by its spectral power density function $S(f)$, which is the Fourier Transform of the autocorrelation function of the signal, as stated in the Wiener-Khintchine theorem \cite{Wie30,Khi34}. We can use a specialized result for periodic random pulses to give analytic formulas for the functions $S(f)$ in different cases of interest. 

We consider periodic pulses with a random width, $w$. We can give each period an integer index, $i$ and call $w_i$ to the realization of this random variable in period $i$.

For these periodic pulses, the power spectrum can be written in terms of the Fourier Transform $U(f)$ of the basic pulse $u(t)$ that is an individual pulse at the origin. Both $u(t)$ and $S(f)$ are parametric functions that depend on the value of $w$. The power spectrum $S(f)$ is determined by the expected values $\Expect{|U(f)|^2}$ and $|\Expect{U(f)}|^2$ for the concrete distribution of $w$. Following \cite{SVG95, Mid60}, we can show the unilateral spectrum
\begin{equation}
S(f)=S_{cont}(f)+S_{disc}(f)
\end{equation}
has two parts, a continuous spectrum
\begin{equation}
S_{cont}(f)=\frac{2}{T}\left( \Expect{|U(f)|^2} -|\Expect{U(f)} |^2\right)
\end{equation}
and a discrete spectrum
\begin{equation}
S_{disc}(f)=\frac{2}{T^2} |\Expect{U(f)}|^2 \sum_{k=0}^{\infty} \delta(f-k f_0). 
\end{equation}

The area below $S_{cont}(f)$ gives the average power of the random signal in a given frequency band. The value of the terms $\frac{2}{T^2} |\Expect{U(kf_0)}|^2$ of the discrete spectrum give the total power for the $k$th harmonic. 

The pulse width modulation scheme of Figure \ref{QPWM} can be analysed with this model. In each period, we have a width $w_i$ that depends on the found photon number $n$, which follows a Poisson distribution. In the calculation, it is useful to find the expected value of a function $e^{jna}$ for a constant $a$ and a Poissonian random variable $n$ with mean $\lambda_T$. The value is the characteristic function of the random variable $n$ at point $a$ and is given by
\begin{multline}
\Expect{e^{j na}}=e^{-\lambda_T}\sum_{n=0}^{\infty}\frac{\lambda_T^n}{n!}e^{jna}=e^{-\lambda_T}\sum_{n=0}^{\infty}\frac{\left(\lambda_Te^{ja}\right)^n}{n!}\\
=e^{-\lambda_T}e^{\lambda_T{e^{ja}}}=e^{\lambda_T(\cos(a)-1)}e^{j\lambda_T \sin(a)}.
\label{ExpExp}
\end{multline}
 
We consider a general random width of the form
\begin{equation} 
w=TD(1-b)+b\frac{n}{\lambda_T}TD,
\label{genwidth}
\end{equation}
which preserves the average duty cycle ($\Expect{w}=TD$) when $\Expect{n}=\lambda_T$. A width of this form can be achieved with a reference constant voltage plus the random voltage at the detector, which is proportional to the number of photons. Depending on the amplitude of the sawtooth function different gain factors might be in order. The width includes a modulation parameter $b$ that allows us to adjust the variability of the width in case there are further functional requirements. 

In a PWM modulation where each $w_i$ is a realization of a random variable of the form in (\ref{genwidth}), with a Poissonian $n$, for square pulses of amplitude $A$, 
\begin{equation}
U(f)=\int_{0}^{w_i}A e^{-j 2\pi f t} dt=\frac{A}{-j 2 \pi f}\left( e^{-j2 \pi f w_i}-1\right).
\end{equation}
We can write the relevant expectation values in terms of the functions
\begin{equation}
E_{\lambda_T}(x)=e^{\lambda_T\left(\cos\left(2\pi x \frac{b T D }{\lambda_T}\!\right)-1\right)}
\end{equation}
and
\begin{equation}
C_{\lambda_T}(x)\!=\!\cos\!\left(\!2 \pi x T D(1-b)+\lambda_T \sin\left(2\pi x\frac{b TD}{\lambda_T}\!\right)\!\right).
\end{equation}

From the result of equation (\ref{ExpExp}) we can find
\begin{equation}
\label{scont}
S_{cont}(f)\!=\!\frac{A^2}{2 T \pi^2 f^2}\!\left[ 1-E_{\lambda_T}^2\!(f) \right]
\end{equation}
and that the harmonics corresponding to the discrete spectrum have an amplitude
\begin{equation}
\label{sdisc}
S_{disc}(kf_0)\!=\!\frac{A^2}{2\pi^2 k^2}\left[E_{\lambda_T}^2\!(kf_0)\!-\!2E_{\lambda_T}\!(kf_0)C_{\lambda_T}\!(kf_0)\!+\!1\right]. 
\end{equation}
We can check the solution numerically. The spectral density can be estimated with different approximate methods from a set of samples. The presented numerical estimates come from the average of the periodograms of 16 non-overlapping segments following Welch's method \cite{Wel67}. Each power spectral density has been estimated from a sample of 1000 random pulses of a signal with a period $T=1$ ms ($f_0=1$ kHz) and different amplitudes, $A$, and duty cycles $D$. The program uses a sampling frequency $f_s=1$ MHz and a Hanning window. 

\begin{figure}[ht!]
\includegraphics{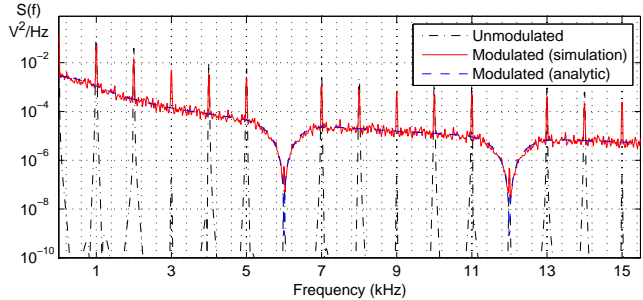}
\caption{\label{Scontb1l2D13}Spectrum of a quantum modulated pulse with amplitude $A=5$, $T=1$ ms ($f_0=1$ kHz), $\lambda_T=2$, $b=1$ and duty cycle $D=\frac{1}{3}$.}
\end{figure}

\begin{figure}[ht!]
\includegraphics{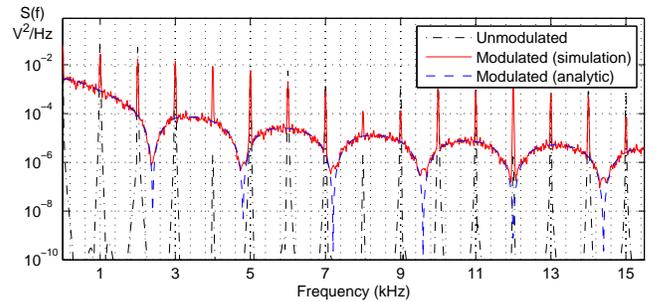}
\caption{\label{Scontb05l03D14} Spectrum of a quantum modulated pulse with amplitude $A=5$, $T=1$ ms ($f_0=1$ kHz), $\lambda_T=0.3$, $b=0.5$ and duty cycle $D=\frac{1}{4}$.}
\end{figure}

Figures \ref{Scontb1l2D13} and \ref{Scontb05l03D14} show there is a good fit between the analytic and the numerical results for the continuous spectrum. The discrete spectrum has been estimated by subtracting the analytic formula for $S_{cont}(f)$ from the numerical results for $S(f)$ and then integrating the resulting delta function around each harmonic. Tables \ref{tableb1l2D13} and \ref{tableb05l03D14} show that the theoretical results are also in good agreement with the numerical estimate. 

The Figures also include the estimated averaged periodogram of the periodic unmodulated signal. Although there are exact results for the energy in each harmonic, taking the periodogram allows a better comparison with the estimated spectrum of the random signals, particularly as far as the delta functions are concerned. 

\begin{table}[ht!]
\caption{\label{tableb1l2D13}Discrete spectrum of the signal shown in Figure \ref{Scontb1l2D13}.}
\begin{ruledtabular}
\begin{tabular}{lcccccccc}
 &1&2&3&4&5&6&7&8\\
\hline
Unmodulated            & 3.80&  0.95&  0.00&  0.24&  0.15&  0.00& 0.08& 0.06\\
$S_{disc}$ (analytic)  & 1.59&  0.32&  0.14&  0.08&  0.06&  0.00& 0.03& 0.02\\
$S_{disc}$ (simulation)& 1.50&  0.33&  0.14&  0.09&  0.06&  0.00& 0.03& 0.02\\
\end{tabular}
\end{ruledtabular}
\end{table}

Figure \ref{Scontb1l2D13} and Table \ref{tableb1l2D13} show the first harmonics of the spectrum of an example quantum pulse width modulation system where the source has an average of two photons per period ($\lambda_T=2$). The first harmonics have most of the energy of the signal and are the most important targets when we want to reduce interference. The quantum modulation takes the power from the discrete harmonics and distributes it through the spectrum. There appears a continuous spectrum and there is a general reduction in the amplitude of the discrete harmonics. In Table \ref{tableb1l2D13} we can see the attenuation, which reaches around 3.8 dB in the power of the fundamental harmonic, the harmonic that gives the peak power that must be reduced in order to pass the regulatory tests. There are also some new harmonics at frequencies where the original signal cancels. In the example with $D=\frac{1}{3}$, the $3m f_0$ harmonics for integer $m$ vanish in the original signal, but the structure lost in the random modulation produces them again. Still, the new power distribution has a smaller maximum power. 

\begin{table}[ht!]
\caption{\label{tableb05l03D14}\!Discrete spectrum of the signal shown in Figure \ref{Scontb05l03D14}.}
\begin{ruledtabular}
\begin{tabular}{lcccccccc}
 &1&2&3&4&5&6&7&8\\
\hline
Unmodulated             & 2.53&  1.27&  0.28&  0.00&  0.10&  0.14&  0.05&  0.00\\
$S_{disc}$ (analytic)   & 0.82&  0.41&  0.40&  0.21&  0.16&  0.05&  0.02&  0.00\\  
$S_{disc}$ (simulation) & 0.79&  0.40&  0.40&  0.21&  0.16&  0.05&  0.02&  0.00
\end{tabular}
\end{ruledtabular}
\end{table}

The average number of photons per period can also be smaller than 1. Figure \ref{Scontb05l03D14} and Table \ref{tableb05l03D14} show an example for $\lambda_T=0.3$, $b=0.5$ and $D=\frac{1}{4}$, where there is around 4.9 dB of attenuation in the fundamental harmonic. 

While this kind of quantum modulation does spread the power even for small photon numbers, there are a few precautions worth mentioning. If the voltage at the detector is greater than the amplitude of the sawtooth signal ($V_n> A_s$), the width of the pulse is exactly $T$. There is a threshold number of photons above which the width remains constant. In the proposed modulation system the width probability density follows, in reality, a truncated Poisson distribution. However, for high enough $\lambda_T$ and moderate $D$s, the probability of detecting a number of photons so high that they take the detector voltage above the threshold is negligible and the analytic formulas in Equations (\ref{scont}) and (\ref{sdisc}) are a good model. When $\lambda_T$ is small, the modulation parameter becomes more important. Roughly, if $bD$ is sufficiently smaller than $\lambda_T$, it is still unlikely to go beyond the threshold.

Even if the given model is not correct, the modulation system can be simulated and there is power spreading. However, we need to be careful to preserve the desired duty cycle. The expected value $\Expect{w}$ is no longer equal to $DT$. 

\begin{figure}[ht!]
\includegraphics{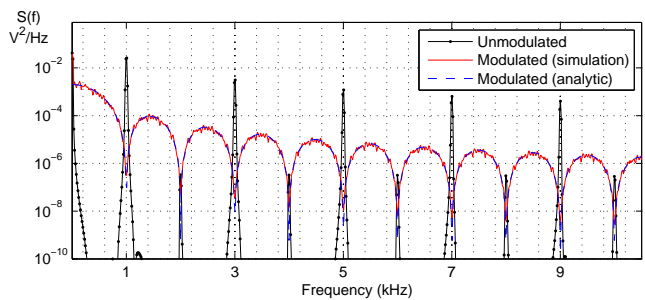}
\caption{\label{onoff}Power spectral density for an on-off quantum modulation when $\lambda_T=-\ln(1-D)$ for $D=0.5$ and pulses with $A=2$. The discrete spectrum vanishes.}
\end{figure}

We can give a simple example if we consider a system with $w_i=anT$, for $a>1$ ($b=1$, $D>\lambda_T$). In practice, this is an on-off modulation in which we have a zero pulse when no photon arrives, with probability $e^{-\lambda_T}$, and a width $T$ when the detector finds one or more photons, with probability $1-e^{-\lambda_T}$. Now
\begin{equation}
S_{cont}(f)=\frac{A^2 e^{-\lambda_T}}{ T \pi^2 f^2}\!\left( 1-e^{-\lambda_T}\right)\left( 1-\cos(2 \pi f T) \right)
\end{equation}
and the discrete spectrum vanishes. The average duty cycle is
\begin{equation}
D=\frac{1}{T}\Expect{w}=(1-e^{-\lambda_T}),
\end{equation}
which can be selected by adjusting the photon source to an intensity such that $\lambda_T=-\ln(1-D)$. Figure \ref{onoff} shows an example of this on-off modulation when we want a duty cycle $D=0.5$. The maximum power outside the DC component has been reduced by around 26 dB.

\section{Discussion}

These few examples show the potential of quantum modulation as a new method for spread spectrum electromagnetic interference reduction. On-off modulation seems particularly efficient and easy to implement. A binary photon detector with no photon counting capabilities suffices. Current avalanche photodiodes can be used. The system can use a part of the period, $T_d$, to detect the photons and reserve the rest for the diode's recovery. The relevant average photon number is then $\lambda T_d$.

The physical implementation of these modulation systems requires only small modifications to existing PWM switching-mode power supplies. The quantum light source and detector could even be integrated side by side in the same piece of semiconductor. A technical aspect that has been left aside for clarity is the presence of noise. Gaussian white noise is likely to superpose to the voltage at the detector and modify the width of the pulses. For a controlled amount of noise, where the quantum signal still dominates, the global effect can be easily simulated and shows only small deviations from the predicted spectrum. 

The proposed width modulation systems are just an example of the possibilities of quantum modulation. There are also interesting results for quantum pulse position modulation and modulations based on the time of arrival of two consecutive photons that will be discussed in a separate paper. 

\newcommand{\noopsort}[1]{} \newcommand{\printfirst}[2]{#1}
  \newcommand{\singleletter}[1]{#1} \newcommand{\switchargs}[2]{#2#1}
\end{document}